# EMITTANCE AND PHASE SPACE TOMOGRAPHY FOR THE FERMILAB LINAC*

C. D. Moore[!], F. Garcia, T. Kobilarcik, G. Koizumi, C. Johnstone, D. Newhart, FNAL, Batavia, IL 60510 U.S.A.

*Abstract*
The Fermilab Linac delivers a variable intensity, 400-MeV beam to the The MuCool Test Area[1] experimental hall via a beam line specifically designed to facilitate measurements of the Linac beam emittance and properties. A 10 m, dispersion-free and magnet-free straight utilizes an upstream quadrupole focusing triplet in combination with the necessary in-straight beam diagnostics to fully characterize the transverse beam properties. Since the Linac does not produce a strictly elliptical phase space, tomography must be performed on the profile data to retrieve the actual particle distribution in phase space. This is achieved by rotating the phase space distribution using different waist focusing conditions of the upstream triplet and performing a de-convolution of the profile data. Preliminary measurements using this diagnostic section are reported here.

## INTRODUCTION

The MTA beam line has been specifically designed to facilitate measurements of the Fermilab Linac beam emittance and properties utilizing a long, 10m, element-free straight. Linac beam is extracted downstream of the 400-MeV electrostatic chopper located in the Booster injection line with the entire Linac beam pulse directed into the MTA beamline. Pulse length manipulation is provided by the 750-keV electrostatic chopper at the upstream end of the Linac and, using this device, beam can be delivered from 8 μsec up to the full 50 μsec capability of the Linac.

The 10 m diagnostic straight both exploits and begins at the 12' shield wall that separates the MTA Experimental Hall and beamline stub from the Linac enclosure. Since three profile measurements completely determine the Courant-Snyder parameters in a straight, multiwires have been installed at the upstream, center and downstream locations to provide the required three profiles. The first profile monitor has been installed upstream of the shield wall, and 5 m upstream of the central, or focal-point monitor. The final one is 4.3 m downstream. The locations of the emittance measurement diagnostics are shown in Figure 1 (top).

A small, approximate beam waist located near a center profile monitor reduces the number of unknown linear optical parameters to two Courant-Snyder parameters, β and ε, since, α, or the rotation of the phase ellipse can be determined by propagating the beam envelope from this waist (using the simple linear transfer matrix that describes a drift). The optics are designed to generate a waist on approximately positioned at the center monitor, MW5. (However, three profile monitors completely determine the Courant-Snyder parameters and thus provide a check for assumptions of a local waist.) In addition three monitors are necessary for a more detailed analysis; i.e. phase-space tomography, in the event of a non-elliptical beam. The small number of variables and the large change in beam size (with no intervening magnetic elements) reduce the systematic uncertainties and errors associated with the measurement. Thus the long magnet-free straight enables a virtually systematic-free measurement of Linac beam properties, in particular emittance.

## BEAMLINE OPERATION

The MuCool beamline must operate parasitically to the Fermilab HEP program. Beam is therefore fully extracted on a single 15 Hz tick which corresponds to the maximum duty cycle of the Fermilab Linac. Intensity in the MTA beamline is controlled by changing the repetition rate (up to 15 Hz) of a fast extraction C magnet in combination with an electrostatic beam chopper, which can vary the Linac pulse length between 8 and 50 μsec. This corresponds to a pulse intensity of 0.64 - 1.6x10$^{13}$ protons.

Quadrupole-triplet telescopes on either side of the straight form an intermediate waist and further allow variable phase advance across the straight thus providing a flexible and powerful basis for beam tomography. The quadrupole triplet installed upstream of the shield wall focuses the large, 1.5-2" (~95% width) beam through the shield wall onto the center profile monitor located at the exit of the shielding. With the triplet, a small, 0.2-0.5" spot size was produced for the initial measurement reported here on the center monitor. (The multiwires installed, MW4 – MW6, have wire pitches of 2 mm, 0,5 mm, and 1mm, respectively in both transverse planes). The set of optics for emittance measurements is shown in Figure 1 (bottom) for the entire beamline.

## PHASE SPACE AND EMITTANCE

Since three parameters are necessary to describe an elliptical phase space, nominally three profile measurements are required. The most systematic-free measurement of the phase space of a beam is provided by three profile measurements at three different locations in a drift; a drift that is sufficiently long to capture a significant change in measured beam profiles. Measured profiles must be consistent with the resolution and active area of the profile monitor in order to determine accurately the profile width - the only physical parameter which is directly measured. The optimal drift is correlated to the beam emittance and the upstream focusing optics with the goal to produce a large difference in the measured profiles on different monitors. Beam divergence is the most difficult to measure and is a constant in a drift, but cannot be determined accurately for small changes in beam widths. The smaller the emittance, or the weaker the focusing in the drift, the longer the drift required to effect a measurably significant change in beam size due to smaller beam divergence.

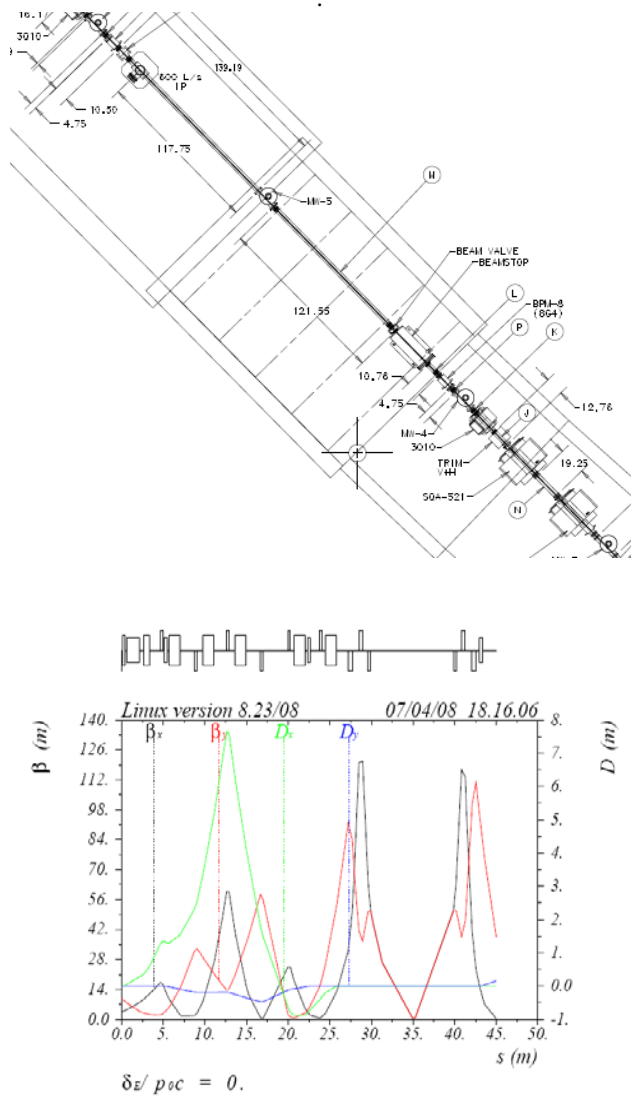

Figure 1. The long straight section instrumented for beam tomography (top) and the emittance mode lattice (bottom).

However, if an upstream focusing system – such as a quadrupole telescope – is applied then the optics can be adjusted across a suitable drift such that a either 1) a waist or 2) a minimum spot size is established at one monitor, and then only one additional monitor is required to complete a phase space and emittance measurement. This provides additional checks on the 3-profile results – particularly in the event that data from one of the profile monitors are noisy or suspect. The following discusses the measurement of emittance for an arbitrary but elliptical beam phase space as measured in a drift using 3 profile monitors and then the special case in which only two profile monitors are required.

## EMITTANCE MEASUREMENTS

An absolute determination of the beam phase space) can be effected by 3 profile monitors in a drift with no assumptions on beam properties outside of an invariant ellipse. The following (Method 1) is an analytical derivation[2] using a Courant-Snyder parameterization of the beam envelope equation.

$$\epsilon = \frac{\sigma_1^2}{S_1}\sqrt{\left(\sigma_2/\sigma_1\right)^2 - C_1^2 + C_1 S_1 \xi - S_1^2 \xi^2/4} \cdot \pi\, mm - mr$$

Results from the 3-monitor analytical solution can be corroborated using any optics program such as MAD and "fitting" the optics to the profile widths (Method 2). The results should be identical in the case of 3 monitors. When using a fitting routine, and given errors in the profile width measurements, additional monitors would better constrain the phase space solution. For the case of more than 3 monitors, a least-squares fit can be readily performed using MAD, for example.

A two-profile method (Method 3) can be applied when a waist or near-waist condition is achieved at one of the monitors. The minimum beam size in a drift always coincides with a beam waist or upright ellipse ($r_{12}$, $\alpha=0$). However, for a fixed beam profile monitor, the minimum beam size as measured at a profile monitor does not correspond to a waist; the waist occurs upstream of the minimum beam size simply because stronger focusing (a shorter focal length) produces a smaller *measured* spot size at the profile monitor. For a long straight, the difference between the waist and the minimum spot size at the 2$^{nd}$ detector is insignificant insofar as the measurement of emittance is concerned. For two monitors spaced relatively equidistant from a central monitor, waist/symmetric conditions can also be verified by "equal" profile measurements. Although the present monitor configuration is ± 5.001/4.312 m ($L_1$ and $L_2$, respectively) about the central one, the relationship between the distance and beta function relative to a waist (assumed to coincide with the central monitor) provides a quick check of on the assumption of near-waist conditions at the central monitor.

The relationship of the waist to the minimum spot size can be fully derived. This relationship is useful in that it demonstrates that only two profile monitors will still supply an accurate emittance – the errors associated with the emittance remain dominated by determination of profile widths and particularly in the case of a non-

elliptical phase space. For a waist at a center monitor, #2, the equation for emittance[1] becomes:

$$\varepsilon_{rms} = \frac{\sigma_2^2}{\beta_2} = \frac{\sigma_2^2}{L_2}\sqrt{\left[\frac{\sigma_3^2}{\sigma_2^2} - 1 - \frac{2L_2}{L_1} - \left(\frac{L_2}{L_1}\right)^2\right]}$$ or

$$\varepsilon = \frac{\sigma_2 \sigma_1}{L_1},$$

Depending on whether the downstream (#3) or upstream (#1) profile is applied[2].

## RESULTS

The Fermilab Linac beam deviates significantly from a Gaussian, exhibiting a more triangular shape, and therefore a weighted mean ($\mu$) and rms ($\sigma$) is calculated for channels above background (n=channel number) using the absolute value for the signal, $|P(n)|$: $\mu = \frac{\sum_n n|P(n)|}{\sum_n |P(n)|}$ and $\sigma_{rms}^2 = \frac{\sum_n (n-\mu)^2 |P(n)|}{\sum_n |P(n)|}$. An approximate 95% (3 x rms) point is chosen as the cutoff for the channels contributing to the rms calculation. Background and noisy wires beyond the signal area cause significant error in the rms calculation. (A constant threshold is not subtracted because it makes an insignificant difference in the rms value.) Raw data from MW5 is depicted in Figure 2. Table 1 gives the results of the peak and rms values for each distribution. Table 2 summarizes the results using the 3 approaches to calculating emittance. The small difference between the 3-monitor analytical result compared with the MAD fit is simply the accuracy (decimal place) of the profile width used as input to MAD.

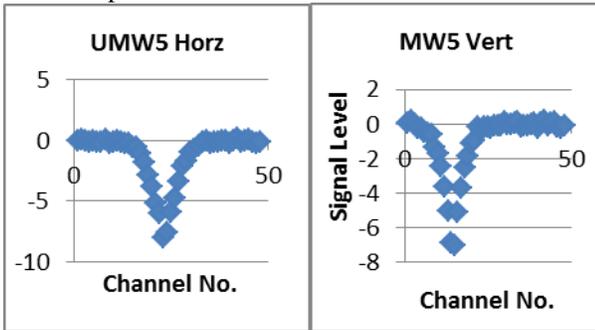

Figure 2. Raw wire profile data.

Table 1. Mean and rms values from the raw profile data.

| Wire | Peak (Wire #) | RMS (# of channels) | RMS x pitch (mm) |
|---|---|---|---|
| Vert: UMW4 | 18.58 | 3.11 | 6.23 |
| UMW5 | 14.52 | 3.34 | 1.67 |
| UMW6 | 32.80 | 6.88 | 6.88 |
| Horz: UMW4 | 33.59 | 2.20 | 4.40 |
| UMW5 | 23.43 | 3.36 | 1.68 |
| UMW6 | 10.32 | 4.14 | 4.14 |

Derivation of the corresponding Courant-Snyder parameters[2] provides additional information about the waist assumption or proximity to the waist. The different methods yield the following Courant Snyder function of Table 3.

Table 2. Results for emittance calculations.

| Emittance | Method 1 $\pi$ mm-mr | Method 2 (MAD fit) | Method 3 Wires 1&2 (2&3) |
|---|---|---|---|
| $\varepsilon_y$ | 2.00 | 1.98 | 2.08 (2.37) |
| $\varepsilon_x$ | 1.45 | 1.39 | 1.48 (1.05) |

Table 3. Derived Courant-Snyder functions for the different methods.

| Courant Synder functions | Method 1 | Method 2 (MAD fit) | Method 3 Wires 1&2 |
|---|---|---|---|
| MW4: $\beta_y$, $\alpha_y$ | 19.34 m, 4.14 | 19.58 m, 4.13 | 18.65 m, 4.27 |
| $\beta_x$, $\alpha_x$ | 13.33 m, 2.36 | 13.93 m, 2.43 | 13.10 3.07 |
| MW5: $\beta_y$, $\alpha_y$ | 1.39 m -0.55 | 1.41 m, -0.54 | 1.34 m -0.27 |
| $\beta_x$, $\alpha_x$ | 1.94 m -0.11 | 2.03 m, -0.09 | 1.91 m, -0.38 |
| MW6: $\beta_y$, $\alpha_y$ | 23.59 m -4.60 | 23.88 m -4.58 | 22.76 m -3.72 |
| $\beta_x$, $\alpha_x$ | 11.78 m -2.35 | 12.33 m -2.26 | 11.58 m -2.97 |

## SUMMARY

These data represent a first-pass measurement of the Linac emittance based on various techniques. It is clear that the most accurate representation of the emittance is given by the 3-profile approach. Future work will entail minimizing the beam spot size on MW5 to test and possibly improve the accuracy of the 2-profile approach. The 95% emittance is ~18$\pi$ in the vertical and ~13$\pi$ in the horizontal, which is especially larger than anticipated – 8-10$\pi$ was expected. One possible explanation is that the entire Linac pulse is extracted into the MTA beamline and during the first few microseconds, the feed forward and RF regulation are not stable. This may result in a larger net emittance observed versus beam injected into Booster, where the leading part of the Linac beam pulse is chopped. Future studies will clearly entail a measurement of the emittance vs. pulse length.

One additional concern is that the Linac phase space is most likely aperture-defined and non-elliptical in nature. A non-elliptical phase-space determination would require a more elaborate analysis and provide another explanation of the large emittance measured.